\documentstyle[]{IEEEtran}

\begin{document}
\title{%
Summary and Outlook of the International Workshop on Aging Phenomena in Gaseous Detectors
(DESY, Hamburg, October, 2001)}
 
\author{ 
M. Titov$^*$\footnote{ $^*$Featured talk at the IEEE Nuclear Science Symposium
and Medical Imaging Conference, November 4-10, 2001, San Diego, USA},
M. Hohlmann, C. Padilla, N. Tesch}

\maketitle
 
\begin{abstract}

High Energy Physics experiments are currently entering a new era which requires
the operation of gaseous particle detectors at unprecedented high rates and
integrated particle fluxes. Full functionality of such detectors over the lifetime of an
experiment in a harsh radiation environment is of prime concern to the involved
experimenters. 
New classes of gaseous detectors such as large-scale straw-type detectors, 
Micro-pattern Gas Detectors and related detector types
with their own specific aging effects have evolved since the first
workshop on wire chamber aging was held at LBL, Berkeley in 1986. 
In light of these developments and as detector aging is a notoriously complex field,
the goal of the workshop was to provide a forum for interested experimentalists to review
the progress in understanding of aging effects and to exchange recent experiences.

A brief summary of the main results and experiences reported at the 
2001 workshop is presented,
with the goal of providing a systematic review of aging effects in state-of-the-art and future gaseous detectors.

\end{abstract}

\section{ Introduction }

 Aging effects in proportional wire chambers, a permanent degradation of 
operating characteristics under sustained irradiation, has been and still remains the
main limitation to their use in high-rate experiments~\cite{charpak}. 
Although the basic phenomenology of the aging process has been 
described in an impressive variety of experimental data, it is nevertheless
difficult to understand any present aging measurement at a microscopic level and/or to
extrapolate it to other operating conditions. 
 Many chemical processes are expected to occur simultaneously 
in the gaseous discharges surrounding the wire, and consequently a 
quantitative description of aging effects, which would require 
as a minimum a detailed analysis of all gas-phase and gas-surface 
reaction products, is currently not available.
There is much experimental information, well summarized 
in~\cite{vavra}-\cite{cern1}, which suggests that wire chamber lifetime
may be extremely sensitive to the nature and purity of
the gas mixture, different additives and trace contaminants, materials
used in contact with the gas, geometry of electrodes and configuration of electric field. 
 The 'classical aging effects', well known since the 
advent of wire chambers, lead to the formation of 
deposits, conductive or insulating, on the electrode surfaces
and manifest themselves as a decrease
of the gas gain due to the
modification of electric field,
excessive currents, self-sustained discharges, or sparking.
Traditionally, the aging rate has been  parameterized as a normalized gas gain 
loss: $R = - \frac{1}{G} \frac{dG}{dQ} ( \%~per~C/cm )$,
where $G$ is the initial gas gain, 
$dG$ is the loss of gas gain after collected charge $dQ$ per
unit length~\cite{kadyk}.
 However, the assumption 
that the aging rate is only a function of the total accumulated charge
has not been confirmed for gaseous detectors operated in high-rate environments.
 In reality, the rate of polymer formation 
depends upon many microscopic variables such as 
cross-sections of electron and photon processes and their energy distributions in gas avalanches,
molecular dissociation energies, as well as densities of electrons, ions
and free radicals. 
Consequently, one may expect that the aging rate 
could also be affected by macroscopic parameters, such as
gas gain, ionization density and radiation intensity, 
which are directly related to the basic microscopic variables. 
Several results presented at the 2001 workshop
clearly indicate that such dependencies do exist.

Some of the conclusions from the 1986 workshop are still valid in 2001. However,
the dramatic increase in charge (up to~1.0~$\frac{C}{cm~wire}$ per year),
which is expected to be accumulated on sensing electrodes in the new high rate experiments,
poses much more stringent constraints on the radiation hardness of materials and gas mixtures, 
assembly procedures, and basic rules for construction and operation of gaseous detectors,
than previously encountered. 
Only a limited choice of gases have been demonstrated to tolerate such doses.
Moreover, recent experience with straws and honeycomb drift tubes revealed
that chemical etching processes leading to a dramatic damage 
of gold-plating on wires could occur in non-polymerizing $CF_4$ mixtures 
at exceedingly high current densities.
 These new developments since the 1986 workshop
raise a question about the adequacy of using $CF_4$-based mixtures for long-term, high-rate applications.

 The scientific program of the 2001 workshop 
addressed specific questions which, as reported by many authors, are of a
primary interest: classical aging effects, 
models and insights from plasma chemistry, 
materials for detectors and gas systems, 
lessons learned from detector operation at high radiation intensities,
new aging effects,
experiences with large systems and recommendations for future detectors.
 About 100 detector experts attended the 4 day workshop,
and 10 invited talks, 31 contributed talks and 9 posters were presented
in 7 sessions~\cite{workshop_proceedings}.

\vspace{0.5cm}

\section {General Characteristics of Aging Processes}
 
\subsection{Classical Aging Effects}

The 'classical' aging effects are the result of chemical reactions
occurring in avalanche plasmas near anodes in wire chambers 
leading to formation of deposits on electrode surfaces.
During gas avalanches many molecules break up 
in collisions with electrons, de-excitation of atoms, 
and UV-photon absorption processes. 
Whereas most ionization processes require electron energies 
greater than 10~eV,
the breaking of covalent molecular bonds and formation of free
radicals requires only 3-4~eV, and can lead to a higher concentration of 
free radicals than that of ions in the  gaseous discharges.
Consequently, free-radical polymerization 
is regarded as the dominating mechanism of wire chamber aging.
 Since free radicals are chemically very active they will either
recombine to form the original molecules or other volatile species,
or may start to form new cross-linked molecular structures
of increasing molecular weight.
 When the polymerized chain becomes large enough
for condensation to occur, it will diffuse to an
electrode surface. 

It is worthwhile to mention that one has to distinguish 
between formation of polymers in the gas avalanche near the anode wire
and their deposition on electrode (anode or cathode) surfaces.
The polymer deposition mechanism 
can be viewed as a phenomenon that occurs whenever the gaseous 
species fails to bounce back after a 
collision with an electrode surface, including a surface layer of molecules
previously formed in the gas discharges. 
Initially, the polymer could be attached to the surface very weakly,
unless some additional chemical reactions 
take place between the polymer atoms and atoms of the wire material.
Moreover, many free radicals are expected to have 
permanent or induced dipole moments so that
electrostatic attraction to a wire can also
play a significant role in the polymer deposition process.
 For the inert gold-plated anode wires the probability for polymers to stick
to the surface is rather small until the creation of the first monolayer of deposits, 
which may significantly increase further deposition. 
The influence of surface wire quality on anode aging and a model of polymer film
growth is proposed in~\cite{blinov}.
 The importance of reactions between the electrode material and 
polymers produced in avalanches for the deposition mechanism
can be illustrated by the following examples:
\begin{itemize}
\item Non-gold anode wires react with fluorine radicals produced in an avalanche
to form resistive metal fluorides.
Many studies have demonstrated 
excellent aging properties, up to 10~$\frac{C}{cm \cdot wire}$,
of $CF_4/i C_4 H_{10}$ (80:20) gas avalanches~\cite{opensh1}-\cite{kadyk2}, which also has the ability
to etch silicon-based and hydrocarbon deposits on previously
aged gold-plated wires~\cite{opensh3,chemmod}. However, extensive deposition was observed
on unplated wires irradiated in $CF_4/iC_4H_{10}$ (80:20)~\cite{chemmod,dissert}. 
\item Exceedingly large aging rates were observed in pure $CF_4$ and in
$Ar/CF_4/O_2$ (50:40:10)~\cite{chemmod}-\cite{vavra3}, which are  typical
etching gases and reluctant to polymerize.
 This effect was related to the chemical processes at the cathode,
where trace fluorocarbon deposits were found resulting in a loss of gas gain and not in a self-sustained Malter discharge. 
\end{itemize}

The self-sustained discharge (Malter effect)~\cite{malter}, which is due to a thin 
insulating layer deposited on a conducting cathode by a polymerization mechanism, 
is one of the most devastating phenomena of all aging effects. 
 The resistivity of the insulating layer 
defines the maximum rate capability of the detector before 
the onset of field-emission of electrons from the cathode,
which starts if the rate of ionic charge 
neutralization across the dielectric film is smaller than the rate of 
ion charge build-up~\cite{vavra2,bohm}.
 There exists evidence that certain metal oxide coatings
on the cathode and/or simply the cathode material itself (eg. carbon-loaded polycarbonate foil)
may not be conducting enough and could cause Malter-like breakdowns
in the presence of large localized ionization densities~\cite{kadyk,bohm,padilla}.
 Several other factors may facilitate its ignition, such as 
highly ionizing particles, sparks, 
sharp points on electrodes causing corona discharges, or thin anode wires~\cite{vavra2}. 
 It is easy to ignite Malter currents in a detector operating with hydrocarbon
gases at elevated high voltages~\cite{boyarski} 
or forcing chambers to breakdown~\cite{foster}-\cite{sadrozinski}, and in a 
detector, which has been previously exposed to TMAE gas~\cite{vavtmae1}.
The CRID RICH detector~\cite{abe} with an excellent 3-dimensional single electron
reconstruction capability allowed the first imaging of the onset of the Malter effect,
which starts from sporadic bursts of single electrons
from a localized cathode spot~\cite{vavtmae1}. 
 Such a positive feedback between 
electron emission at the cathode and anode amplification will lead
to high ionization densities at distinct chamber locations. 
This, in turn, can initiate the production of new
reactive species at much larger rates, thus 
promoting more deposits to form at the same cathode spot 
to an extent sufficient to establish a classical self-sustained Malter discharge.
The most dangerous consequence of this phenomena is that the Malter effect
could easily spread over a large area, if it goes undetected for a long period of operation,
thus causing irreparable damages to the chamber.

Many experiments have demonstrated that the addition of $H_2O$ or alcohols -- after the insulating layer at the cathode is already formed -- tends to stabilize the detector
operation, but not to cure the Malter effect~\cite{vavra,kadyk}.
When these additives are removed, usually the chamber suffers 
from Malter effect again.
 Recently, it was discovered that the addition of oxygen (0.02-0.05~$\%$) or $CO_2$ (5~$\%$) to the 
damaged chamber, which showed a self-sustained dark current with $He/iC_4H_{10}$ (80:20), could
revert or cure a Malter breakdown in the presence of high current density~\cite{boyarski}.
When the oxygen is removed, the chamber can still operate at a high ionization level 
(although it will start to age again without additive).
It is also worthwhile to mention that the possibility of reanimation of anode wires aged in
hydrocarbon gases
by means of sputtering was demonstrated in  $Ar/O_2$ (99:1) 
and $Ar/CO_2$ (93:7)~\cite{pash,atlas2}. 
These effects support results
from plasma chemistry, where it is known that
oxygen reacts with hydrocarbon molecules and the end products are volatile 
$CO$, $CO_2$, $H_2O$ and $H_2$. 

\subsection{ Wire Chambers vs Plasma Chemistry}

While the specific reactions responsible for wire chamber aging 
are extremely complex, some qualitative approach to the aging phenomena 
in different gases could be obtained
from similarities between chemical processes in plasmas 
of gas avalanches~\cite{vavra,kadyk,chemmod,dissert}
and those that occur in the better-understood low-pressure ($<$~1~Torr)
rf (13.6~MHz) plasmas~\cite{yasuda,boenig}. 
 Although many parameters (electric field, gas pressure, 
electron density, power density)
are vastly different between the two regimes, the electron
energies are not so different. Also, in both
cases the free radicals are most likely the active species involved in polymer formation.

In plasma polymerization,
the overall mechanism of `Competitive Ablation and Polymerization' proves to be a basic 
principle that describes reactions occurring in a plasma polymerization system.
 Considerable fragmentation of the gas molecules or rearrangement
of atoms occurs in the plasma and the extent of the process and the dominating mechanism 
vary with the types of gases and the discharge conditions.
The most important concept here is that both polymer-forming species and species that cause ablation (physical
or chemical etching) of materials are created in the plasma of the original gas.
 The significance of this concept is fully established in perfluorocarbon plasmas, which 
represent the most extreme case of ablation competing with polymer formation.
Actually, $CF_4$-based gases are used for both
etching and deposition processes, the distinction being made by the gas and
its concentration with which $CF_4$ is mixed. 
In general, the addition of oxygenated species shifts the chemistry of $CF_4$
plasmas towards etching, while the addition of hydrogenated species shifts the 
chemistry towards polymerization~\cite{yasuda,yasuda2}. 
In the former case, 
the dissociative products of $CF_4$ and $O_2$ are the most desirable active species
for the etching processes in plasmas~\cite{kushner}-\cite{martz}.
For the latter case, in the absence of hydrogen, products of the
$CF_4$ discharge could act as an effective etching gas especially for $Si$-based deposits, which react
with fluorine to form volatile $SiF_4$. The addition of hydrogen atoms or molecules
to $CF_4$ scavenge $F$ atoms by the formation of more stable $HF$ and produces a mixture
with carbon-enriched ($CF_3, CF_2, CF$) residues.
As the ratio of $F/C$ decreases, perfluorocarbons polymerizes readily, i.e.\ the balance
shifts from ablation to polymerization~\cite{yasuda2,truesdale1}-\cite{motlagh}. 
For instance, very fast polymer formation was observed in $C_6F_6$ 
and $C_2H_2F_4$ plasmas~\cite{yasuda2}.
On the other hand, hydrofluoric acid can chemically attack $HF$-soluble materials
existing in the system.
Under certain conditions, 
$Si$-etching can be accompanied by the 
polymerization of the etching gas $CF_4$ on the $Si$-substrate~\cite{kushner,arikado,winters2}. 

Correspondingly, recent results from wire chamber operation 
also show that both polymerization and etching phenomena can occur 
in $CF_4$-based gases (see section V). 
Particularly, using the same experimental setup, a
lack of apparant aging have been observed in $CF_4/iC_4H_{10}$ (80:20) and 
$CF_4/iC_4H_{10}$ (50:50) mixtures, whereas
heavy carbonaceous deposits were observed on the gold-plated wires in 
$CF_4/iC_4H_{10}$ (95:5), $CF_4/iC_4H_{10}$ (20:80) and $CF_4/C_2H_4$ (95:5) gases~\cite{chemmod}.

 Two other examples, where conclusions from plasma chemistry
are qualitatively applicable to wire chambers are:

\begin{itemize}
\item In plasma chemistry, most organic compounds with oxygen-containing groups 
are generally reluctant to form polymers.
For example, water in plasmas could act as an efficient modifier of the polymer
chain-growth mechanism by reacting with polymer precursors and forming volatile species
(up to 50~$\%$ $H_2O$ was added to the plasma feed gas)~\cite{yasuda}.
In wire chambers, the addition of water (a few hundred to a few thousands $ppm$ of $H_2O$) has been
found to effectively suppress polymerization effects~\cite{kollef,wulf},
to prevent Malter breakdown~\cite{boyarski,kadyk3,sld}, or even to restore the original
operation in aged counters~\cite{algeri,danilov}. 
There is more than one mechanism by which $H_2O$/alcohols can help in wire 
chambers~\cite{vavra2,vavrasl}.
Because of the large dipole moment, these molecules will tend to concentrate near the
electrode surfaces, where polymerization takes place.
Water has an additional advantage in wire chambers since it
increases the conductivity of the partially damaged electrodes - 
a property that can have adverse effects in a MSGC~\cite{cern1}.
\item In plasmas, the characteristic polymerization rate of $Si$ is higher than for $C$~\cite{yasuda2}.
From the viewpoint of wire aging,
even minor traces of $Si$-pollutants in the gas have a much higher tendency
to create deposits, than similar amounts of hydrocarbon molecules.
\end{itemize}

It has to be stated, though, that
 the absence of corresponding systematic studies 
in plasma chemistry with parameters similar to wire chambers 
(atmospheric pressure, power densities, gas mixtures) does not  
allow any quantitative comparisons between the plasma chemistry
and wire chamber processes.

\section{ Experience from laboratory R~$\&$~D experiments.}

Over the last few decades an impressive variety of experimental data
has been accumulated from laboratory tests and detectors installed
at high energy physics facilities. However, there are many contradictory experiences 
obtained in seemingly identical conditions, which means that we do not always
control all parameters that influence aging effects.
It is now well established that -- even if a low aging rate
can be obtained in the laboratory with very pure gas
and otherwise clean conditions -- large-area detectors using the 
same mixture can fail due to severe aging after a relatively small beam exposure.
However, experience from the laboratory, where operating conditions are much better controlled,
can be used to understand some general principles and might help 
to implement these results successfully in large chambers.

There are a lot of experiments 
that clearly indicate premature aging in $Ar/CH_4$ mixtures
exposed to intense 
radiation~\cite{kollef,smith}-\cite{silander}.
Moreover, the aging rate in $Ar/CH_4$ (90:10) 
was found to be mainly a function of current density, i.e.\ the product
of irradiation rate and gas gain, independently from
electrode material and purity of methane~\cite{bouclier,capeans}.
 This observation indicates that $CH_4$ itself polymerizes in the avalanche plasma
due to the hydrogen deficiency of radicals and their 
ability to make bonds with hydrocarbon molecules~\cite{kadyk,yasuda},
and similarly for all hydrocarbon gases.
 Under certain conditions the aging rate in $Ar/C_2H_6$ with alcohol 
can be strongly reduced~\cite{atac,ua1}.
However, noble gas/hydrocarbon
mixtures are not trustworthy for long-term, high-rate experiments.
In order to suppress polymerization of hydrocarbons, oxygen-containing molecules
can be added to the mixture. For $Ar/CH_4/CO_2$, measurements
have shown that sensitivity to aging decreases with decreasing $CH_4$ 
and increasing $CO_2$ content~\cite{pash}.

Dimethylether (DME) appeared in the 1986 workshop 
as a good quencher and a reasonably good radiation-hard gas
for wire chambers operated at high intensities.
The aging rate in DME tends to be lower than the polymerization rate of
ordinary hydrocarbons~\cite{vavra} and several groups reported the absence of 
aging effects in wire chambers
up to large values of accumulated charge~\cite{blinov,bouclier,capeans,openshaw}.
 However, the aging effects in DME appear to be highly sensitive to traces of pollutants
at the ppb level, which are difficult to keep under control in large detectors.
 There is also evidence of high chemical reactivity of DME,
which requires a careful material selection for detector assembly and 
gas system components~\cite{dme}.

Attempts were made to replace organic quenchers with 
aging resistant ones, like $CO_2$. 
  The $Ar(Xe)$/$CO_2$ gases could be in principle absolutely radiation 
resistant under clean conditions;
up to now, there is no well-established mechanism, which could lead
to the formation of anode deposits in these mixtures.
Stable operation up to $\sim 1 \frac{C}{cm~wire}$
was reported for $Ar/CO_2$~\cite{vavra,kadyk,turala,conti} 
and up to $\sim 5 \frac{C}{cm~wire}$ for $Xe/CO_2$ mixtures~\cite{bondarenko}.
However, a gradual decomposition of $CO_2$ can
also occur and the resulting pure carbon can be deposited specifically at the cathode~\cite{pash,co2}.
Sometimes this carbon layer does not affect the performance of drift tubes~\cite{pash}. 
Recent systematical aging tests were performed for the ATLAS muon aluminum 
drift tubes. In order to guarantee reproducibility of the results and 
to study aging behavior under different operating conditions,
26 tubes with $Ar/CO_2$ (93:7)+600~$ppm$ $H_2O$ mixture have been irradiated with an $Am^{241}$ source
up to an accumulated charge of $\sim$~1.3~$\frac{C}{cm~wire}$ 
and 47 tubes with $Ar/CO_2$ (90:10) gas were exposed to a $Cs^{137}$ source
up to $\sim$~0.6~$\frac{C}{cm~wire}$~\cite{atlas2,kollef2,atlas}.
All tubes were 100~$\%$ efficient at the end of these aging runs,
however,
these measurements represented an average performance of the wire over a length of 3~m and were not sensitive to local inefficiencies.
It should be mentioned, that the aging performance of
$Ar/CO_2$ is sensitive to traces of impurities. 
$Si$-based pollutants are one of the sources of aging in $Ar/CO_2$, 
probably due to the production of non-volatile $SiO_2$~\cite{conti,kollef2,adam,gavrilov}.
 Several other experiments also observed aging effects in $Ar/CO_2$,
however, the reasons of the gain reduction
were not identified~\cite{blinov,atlas2,faruqi,kowalski}.

The identification of radicals and fragments formed in the electron avalanches is a means to 
understand and eventually overcome the problems related to the aging of gaseous 
detectors~\cite{kadyk,hess,wise2,vavra5}.
A recent investigation of avalanche products has shown that
seventeen new compounds were identified in the effluent gas stream
from an irradiated proportional counter with $Ar/C_2H_4$ (50:50) mixture~\cite{kurvinen}.
 Some of the observed species (aliphatic hydrocarbons)
contained double or triple bonds which, similar 
to plasma polymerization, can be easily 'opened' in the discharges and polymerize very
aggressively. 
The systematic analysis of light emission spectra in proportional counters may also provide useful
information about basic physics processes in electron avalanches~\cite{sumner,lima}.

\section{ Experience with `standard radiation level' detectors.}

\subsection{ Classical Wire Chambers.}

For a long time, classical wire chambers
of various designs with many thousands of wires have been used
for large-area tracking detectors
in the `standard radiation levels' experiments, i.e.\
with total collected charges $<$~50~mC/cm for the whole period or running.
 Basic rules for the construction and operation of these detectors are known.
Moreover, many of the large wire chambers were built and demonstrated 
to work~\cite{vavra,kadyk,zarubin}.
Nevertheless, recent experience with large systems still shows the appearance  of
aging effects associated mainly with
hydrocarbon polymerization and with presence of pollutants in the gas system.

A time expansion chamber filled with $CO_2/iC_4H_{10}$ (80:20) mixture successfully operated
as a vertex detector of the L3 experiment at LEP. 
After an accumulated charge of $10^{-4} \frac{C}{cm~wire}$ collected during 11 years of running,
there was no sign of aging~\cite{betev}.
Classical aging effects 
(Malter effect or/and sense wire deposits) have been observed 
in the H1 Central Jet Chamber operated with $Ar/C_2H_6$ (50:50) + 0.1~$\%$ $H_2O$ and
in the ZEUS Central Tracking Detector filled
with $Ar/CO_2/C_2H_6$ (83:12:5) + 0.5~$\%$ $C_2H_5OH$ at the HERA $ep$-collider. 
In the former case, the replacement of 0.1~$\%~H_2 O$ with 0.8~$\%$ of ethanol
cured the Malter effect and stabilized the detector operation~\cite{niebuhr}, while
for the latter case the aging problem was alleviated by the addition of $H_2O$~\cite{bailey}.
In both systems there was no clear indication that polymerization is `fed' by the
presence of impurities in the gas system,
indicating that hydrocarbons are the likely source of chamber aging. 

 An abundance of literature exists describing the dramatic effect of certain
gaseous constituents, which may be either due to the
contaminants initially present in the gas system or that result from
outgassing of construction materials upon the aging rate of wire chambers.
 Several examples of large systems, where the presence of pollutants increased
the aging rate many times, have been reported in~\cite{binkley,marshall}.
While laboratory tests with prototype chamber indicated negligible aging rates 
($R < 10~\% /C/cm$),
a much larger aging rate (1000~$\% /C/cm$) was observed
in the large Central Tracking Chamber of the CDF experiment
operated with $Ar/C_2H_6$ (50:50) + 0.1~$\%$ alcohol.
 The analysis of the aged sense wires showed the presence of $C$, $O$ and $Si$ elements.
 After cleaning the gas system components and making changes to reduce aerosols emanating
from an alcohol bubbler, the aging rate was greatly reduced allowing the detector
to operate without dramatic loss in performance.
 The presence of $Si$ is to be pointed out here since silicon has been systematically 
detected in analysis of many wire deposits, 
although in many cases the source of $Si$-pollutant has not been clearly identified.
 The $Si$-compounds are found in many lubricants, adhesives and rubber,
encapsulation compounds, silicon-based grease, various oils, 
G-10, RTV, O-rings, fine dust, gas impurities, polluted gas cylinders,
diffusion pumps, standard flow regulators, molecular sieves, 
and their presence may not necessarily be noted in the manufacturer's documentation~\cite{vavra,kadyk,capeans}.
Because of a high specific polymerization rate 
$Si$ molecules should be avoided in the detector system at all cost.
Consequently, if there is a question whether or not some device may incorporate silicon compounds 
it should be subjected to additional aging test.
Another example is the D0 WAMUS muon drift chambers, which 
suffered fast anode aging when operated with an $Ar/CF_4/CO_2$ (90:6:4) mixture.
Here, the source of the contaminant was outgassing of glass-steel polyester epoxy resin
used in the construction. A cold-trap added to the gas recirculating system 
reduced the aging rate by a factor of ten, while the extreme method of quickly heating the wires just below their melting point (`zapping') succeeded
in blowing hardened sheaths of outgassing products off of the wires; 
thus completely cleaning the aged gold wires `in-situ'~\cite{marshall}.

The chosen examples underline
the importance of having control over all detector parameters, but often
it is quite difficult to draw final conclusions 
since nominally identical detectors connected to the same gas circuit
may perform very differently~\cite{niebuhr,marshall}.
In some cases it might be possible to eliminate harmful impurities
by installing appropriate filters or cold traps in the gas system~\cite{binkley,marshall}.
Many helpful guidelines for construction and operation of classical wire chambers at `low' rates, which have been
compiled over the past 40 years, are summarized in~\cite{schmidt} as follows:

\begin{itemize}
\item Create a moderately clean environment during detector construction and 
clean the gas system components prior to start of operation;
\item Avoid the presence of `bad'
molecules in contact with active gas ($Si$, halogens, sulphur,
plasticizers, outgassing)~\cite{vavra,kadyk};
\item A huge variety of gases can be successfully used 
(noble gases, hydrocarbons, freons, $CO_2$, $DME$, $H_2O$, alcohols,`magic gas',...);
\item Hydrocarbons are the most likely source of aging 
(effect is more pronounced in presence of contamination or
under discharges, sparks, Malter effect). Even with 
addition of water/alcohol the improvements are still limited and it is problematic to consider them for high rate applications;
\item If aging effects are observed despite taking all of the above 
precautions, add suitable additives and/or
identify the source of pollution and clean the gas system.
\end{itemize}

\subsection{ Resistive Plate Chambers.}

 In the 1990's, Resistive Plate Chambers (RPC)
were proposed as an economical and proven
technology ideally suited for large-area detection systems. For example,
both the Belle and BaBar experiments have instrumented their flux returns with RPC's
operated in streamer mode.
However, high chamber currents started to show up in Belle's RPC's almost 
immediately upon installation.
This problem was related to the presence of high levels of water ($\sim 2000 ppm$) in the gas,
which permeated through the walls of the polyethylene tubing.
Operating the RPC's with $Ar/C_4H_{10}/C_2H_2F_4$ (30:8:62) plus water in
streamer mode led to the formation of hydrofluoric acid that etched the 
electrodes made from ordinary glass.
This resulted in the creation of emission points triggering chamber currents.
The problem was finally solved by replacing the plastic tubes with copper ones ($H_2O < 10 ppm$)~\cite{marlow}.

In the BaBar RPC's the electrodes are made of Bakelite coated with linseed oil. 
After an initial period of successful operation with an 
$Ar/C_4H_{10}/C_2H_2F_4$ mixture, 
the RPC's started to show a permanent reduction in efficiency and an increase in dark current. 
The main conclusion of the subsequent extensive $R\&D$ studies
related the BaBar RPC problem to the lack of polymerization of the linseed oil
and formation of oil droplets under the influence of high temperature and 
high currents~\cite{piccolo}.
Further efficiency deterioration mechanisms, that may play an important role in BaBar's RPC's have been proposed by J. Va'vra~\cite{vavra2}. He suggested that
this problem could be due to an electrochemical change of resistivity of fresh linseed oil, modulated by the presence of water in the RPC's.
 A positive example is the L3 RPC at LEP, which 
operated at a very low particle flux 
over 8 years without significant loss of efficiency~\cite{carlino}.
 In contrary to RPC's used in streamer mode at the Belle, BaBar and L3 experiments,
the future LHC experiments will operate RPC's in proportional mode, which is desirable
in terms of total accumulated charge per particle. 
 However, much higher particle fluxes at the LHC require more systematic $R\&D$ studies
of the RPC technology,
since many processes could degrade their performance under high-rate conditions.
Recent results of aging tests for ATLAS, CMS and LHC-B RPC's indicate that
under the right circumstances RPC can withstand large integrated 
doses~\cite{aielli}-\cite{passaleva}.

 It should be stressed though that
the problems with Belle and BaBar RPC's are not `classical aging effects', but
rather unpredictable surface effects, related to the specific choices
 of materials and operating conditions.

\subsection{ Gaseous Photodetectors.}

 Gaseous photon detectors used in high energy physics experiments must ensure
an efficient way of converting $UV$ photons
to electrons with a subsequent detection of single photoelectrons~\cite{ypsilantis,photon}.
For gaseous converters, systematical aging studies have been carried out with
$TMAE$ and $TEA$ vapors~\cite{woody}-\cite{tea},
which are added to the carrier gas to provide photoionization capability~\cite{tmae1}.  
 TMAE is the best material in terms of quantum efficiency, however,
the main obstacle of using $TMAE$ at high rates 
is an exceedingly rapid gas gain 
loss due to deposits on the anode wires~\cite{photon,vavtmae3}-\cite{lau}.
 Several studies have indicated that anode wire deposits can be removed
by heating the wires with elevated currents~\cite{tea,constr};
unfortunately this recovery is followed by a quick drop in gain~\cite{korpar,skrk}.
In addition, all photosensitive materials, and, most probably, their various aging products
are good insulators and may excite self-sustained currents when deposited on 
the cathode~\cite{photon}.
 Systematic studies with $TMAE$ and $TEA$
also allowed to establish basic dependencies between the aging behavior and
dissociation energy or wire diameter~\cite{vavtmae3}:
\begin{itemize}
\item aging rate for $TMAE$ is larger than for $TEA$ ($TMAE$ molecule
is more fragile than $TEA$);
\item aging rate for $TMAE$ and $TEA$ is inversely proportional to the anode
wire diameter;
\end{itemize}

 At low rates, the possibility to use gaseous photon detectors 
on a very large scale at long-term with hydrocarbon/TMAE gases
has been demonstrated for large 4$\pi$ devices (e.g.\ SLD CRID and DELPHI RICH)~\cite{photon}.
 It is worthwhile to mention, that the high reactivity of TMAE with oxygen and other substances
necessitates a very high degree of cleanliness and leak-tightness for gas systems in these detectors.

 In recent years, there has been considerable work in the field of photon
imaging detectors by combining solid photocathodes ($CsI$) and wire chambers or
gaseous electron multipliers. 
 However, several aging tests also revealed degradation of $CsI$ photocathode quantum
efficiency for very high rate environments; a collection of existing aging 
data for these can be 
found in~\cite{photon,vavtmae3,krizan,breskin11}-\cite{dimauro}.

\section{ Aging Experience with High-Rate detectors of the LHC era. }

 The most recent developments in high-energy physics require a
dramatic increase of the radiation intensity encountered by 
gaseous detectors: from $mC/cm/wire$ for the `standard radiation level' detectors
up to $C/cm/wire$ for the new high-rate experiments of the LHC 
era (HERA-B, LHC).
Among the most critical items that affect the lifetime of gaseous detectors
(apart from the gas mixtures) are the materials in contact with gas, assembly 
procedures, gas mixing and distribution systems, and tubing.
 In section III.A we discuss the outgassing properties of several materials and general rules
for assembly of high-rate gaseous detectors and gas systems,  
while sections III.B-D contain the summary of the recent experience with aging problems in gaseous
detectors operated at extremely large particle fluxes.

\subsection{ Choice of Materials for Detectors and Gas Systems}
 
 The increasingly challenging requirements for building and testing the next generation
of large-area gaseous detectors has demanded a concerted effort towards finding
adequate materials for detectors and associated gas systems.
Many non-metallic `good materials' successfully used in the 
`standard radiation level' detectors might nevertheless outgas at a small level,
thus causing fast aging under high rate conditions.
 The lifetime studies of Microstrip Gas Chambers (MSGC) in high intensity environments,
which also had the greatest impact on the understanding of aging phenomena in all types
of gaseous detectors, demonstrate that
the amount of pollutants in the gas system play a major role
in determining the aging properties of the detector~\cite{cern1}. Consequently, the
outgassing from materials, epoxies, joints, tubing, etc. has to be carefully controlled.
 For obvious reasons, the use of glues, plastics and many organic materials is 
unavoidable
in particle detectors. It is therefore very important to choose materials,
which are suitable for the practical mechanical and electrical assembly of a 
gaseous detectors, 
in terms of their possible outgassing effects and radiation robustness.

It is suggested to start by searching for low-outgassing materials in a NASA 
database, which was originally developed for selecting spacecraft materials~\cite{capeans,nasa}.
It contains more than 1600 entries for adhesives, 
500 entries for rubbers and elastomers, 800 entries for potting materials. 
This list can help to pre-select
assembly materials before doing tests matching the specific requirements of each detector.
  
 A large amount of outgassing data for
epoxy compounds, adhesive tapes, leak sealers, rigid materials, O-rings, and 
plastic pipes  
have been accumulated in the framework of the RD-10 project at CERN, which afterwards was merged
with the more specific research on MSGC within the RD-28 project. In the RD-10 tests,
an outgassing box, placed upstream of the strongly irradiated wire counter, 
was used to introduce samples of materials into the gas stream, 
thus allowing a systematic study of outgassing effects on the chamber lifetime. 
Furthermore, the gas flowing from the chamber
was analyzed with a Gas Chromatograph (GC) and Mass Spectrometer (MS) or Electron Capture Device (ECD).
 While for some materials only outgassing properties were verified and 
materials releasing detectable pollutants were rejected, for other `radiation hard materials' 
full evidence of suitability was obtained in long-term aging tests of 
a validated clean detector.
The long list of candidates recommended to be used in the construction of gaseous detectors can be found 
in~\cite{vavra}-\cite{cern1},\cite{bouclier2}-\cite{guarino}
and was summarized at the 2001 workshop~\cite{capeans}. 
The effect of outgassing from materials
on the lifetime of gaseous detector can be illustrated by several examples:
Araldit AW103 epoxy mixed with HY991 hardener did not induce any detectable gas pollutants
in the GC/MS and was also validated by irradiating the wire chamber.
It is presently used as a glue for the construction of ATLAS straws~\cite{guarino} 
and triple-GEM stations of the COMPASS experiment~\cite{compass}. 
The GC/MS analysis of
another popular epoxy, Araldit106 and HV953 U, extensively used in the assembly of older MWPC's,
revealed traces of heavy hydrocarbon molecules in the effluent gas stream, which could be 
partially responsible for the observation of aging reported in~\cite{padilla,kollef2}.
In fact, this glue has shown the largest outgassing rate among all tested glues in~\cite{guarino}.
Interestingly, outgassing
can also be due to an incorrect ratio of hardener to resin or even insufficient
curing time; both factors may largely increase the gas contamination~\cite{capeans,saulir}.

 Gas tubes used for the supply of the active gas have always been the object
of primary attention when analyzing aging effects in wire chambers.
Electro-polished stainless steel and hydrogen-fired copper gas pipes are 
the best choices for gaseous detectors operated in high-rate environments, since they
are free of outgassing and ensure zero gas permeability. However, due to their
high price and concerns for the material budget in the active area of the detector,
many experiments often use cheaper plastic tubes, although these are 
susceptible to outgassing, have high gas permeability and can consequently cause severe aging.
 Particularly, PVC, Teflon and neoprene rubber tubes contain halogen atoms
in molecular chains, which are known to increase drastically the aging rate~\cite{vavra,kadyk}.
Polyethylene tube outgasses water, large alcanes and substituted 
aromatics~\cite{wise2}.  
One of the classical examples, cited by many authors,
shows that the introduction of PVC pipe can initiate a gain reduction, which continues
with the same rate even after the PVC tube is replaced with the original stainless steel tubes~\cite{kotthaus}.
This indicates a potentially very serious problem: one can cause 
permanent damage to a detector by a wrong choice of material
even for a limited period of time. 
Therefore, one should 
use as much steel as possible for gas supply lines, especially in parts 
exposed to high radiation doses.

 Up to now, there is no strong objection to the use of nylon 
(polyamide, RILSAN) tubes if they are not too long.
However, plastic pipes usually introduce water to the gas due to the natural
outgassing and/or due to the diffusion of air humidity through the walls:
as much as $\sim$~1700~$ppm$ of water can be added to the gas
by placing 20~$m$ of nylon pipe at the chamber inlet~\cite{capeans}.
Particularly, nylon tubes were used to introduce water indirectly to the chamber for curing Malter breakdown~\cite{kadyk3}.
 However, the presence of water can cause `bad' surface chemistry as described
above and is therefore
extremely dangerous for certain RPC and MSGC detectors.

The general recommendations concerning the choice of assembly materials and 
rules for the mechanical construction of high-rate detectors, which includes 
adequate assembly procedures, personnel training, quality checks, final testing
as part of fighting against the aging, have been reported at the workshop in~\cite{capeans,schmidt}.
There are clearly many `bad' and a lot of usable materials. However,
a specific material is either adequate or not for a particular detector type and 
operating conditions - one has to do tests matching the specific requirements
of the experiment.
Finally, no spontaneously-chosen materials should be installed in the detector
or gas system in the last minute, before the start of real operation.

\subsection{ Micro-pattern Gas Detectors.}

Future high-luminosity experiments have prompted a series of inventions
of new high-rate gaseous detectors: MSGC, 
MICROMEGAS, Gas Electron Multipliers (GEM) and many others~\cite{proc1,proc2}.  
The systematic research of the physical parameters 
used to manufacture and operate MSGC's, such as substrate and assembly materials, metal 
for the strips, and type and purity of the gas mixtures 
play a dominant role in determining their long-term stability~\cite{cern1,constr1,sharma}.
Despite the  promising performance of MSGC's (high-rate capability, good space accuracy, and excellent multi-track resolution),
there are several major processes, particularly at high rates, leading
to operating instabilities: charge-up of substrates, destructive
microdischarges, and surface deposition of polymers~\cite{cern1}.

The influence of glass conductivity has been verified for MSGC's:
the use of borosilicate glass as a substrate results in rate-dependent modification
of gain due to the radiation induced variation of surface resistivity. Use of
electron-conducting or diamond-coated glass 
solves the problems of short and long-term instabilities for detectors made
on insulating support~\cite{cern1}.
 The problem of microdischarges, induced by heavily ionizing particles
and destroying the electrode structure, turned out to be a major limitation
to all single-stage micro-pattern detectors in hadronic beams~\cite{itr,bressan10}.
The nature and resistivity of the strip material affects the developments of sparks~\cite{bschmidt}.
Aluminum electrodes are more robust against gas discharges than gold.
However, the use of $Al$ electrodes led to the appearance of 
`bubbles' or `craters' on the cathode strips even at modest 
collected charges, while no aging effects
were observed with strips made from gold~\cite{hildebrandt}.
 
Microstrip chambers have been operated with a large variety of gases; 
to prevent fast aging at high rates, convincing evidence suggests again to avoid hydrocarbons in the gas~\cite{msgc1,msgc2}. 
Under optimal laboratory conditions,
absence of any degradation of MSGC performance with $Ar/DME$ has been demonstrated by many groups
up to large accumulated charges~\cite{cern1,sharma,msgc2}-\cite{msgc4}.
However, MSGC-GEM detectors operated with $Ar/DME$ shows fast aging under $X$-rays, 
if the size of the irradiated area is large 
enough, while identical chambers with $Ar/CO_2$ showed no aging~\cite{hildebrandt,hott}. 
 Long-term survival without degradation has been also observed for 
triple-GEM and MICROMEGAS-GEM detectors,
operated with $Ar/CO_2$~\cite{kappler,miyamoto}.
 Unfortunately, $Ar/CO_2$ mixtures have worse quenching properties and are
 more prone to discharges than $Ar/DME$. Protection against sparking
can be significantly improved by adding a small amount ($\sim 0.3~\%$) of 
$H_2O$ to the MSGC. However, in contrary to wire chambers, such an 
addition of water 
led in one case to massive coating on the anode strips both in 
$Ar/DME$ and $Ar/CO_2$ mixtures~\cite{cern1,hildebrandt}.
 These observations underline the importance of careful selection of 
materials and gas mixtures for high-rate applications
and of treating micro-pattern detectors as delicate
devices during production and running phases.

\subsection{ Choice of Gas Mixtures.}

Future high energy and luminosity experiments pose a new challenge for 
gas mixtures, raising the requirement for their radiation
hardness up to $\sim 1 \frac{C}{cm~wire}$ per year.  
Under these constraints only a limited choice of gases is available, and 
from the `conventional mixtures'
only $Ar(Xe)/CO_2$ is demonstrated to tolerate such doses.
 Unfortunately, these mixtures are quite transparent for photons and 
have a low electron drift velocity, which limits their possible application
for high-rate detectors and large drift distances.

About twenty years ago, $CF_4$ was proposed 
as the most attractive candidate 
for high-rate environments~\cite{christ1}-\cite{christ3}.
This is primarily due to the high-drift velocity, high primary ionization,
low electron diffusion and resistance to aging~\cite{vavra3,b1shmidt,fischer}.
Within the broad spectrum of gas mixtures, there is no gas mixture without $CF_4$
that is able to tolerate doses $\sim~10~\frac{C}{cm~wire}$.
It is believed that, when $CF_4$ dissociates in the gaseous discharges
into highly reactive $CF_x$ and $F$ radicals, the atomic fluorine
is very effective in suppressing polymerization effects.
However, $Ar(Xe)$/$CF_4$ mixtures have rather poor energy and spatial resolution due to 
the dissociative electron attachment processes in $CF_4$~\cite{datskos}-~\cite{biagi3}.
Moreover, the $CF_4$ molecule has a small quenching cross-sections of metastable $Ar$-states~\cite{velazco}
and excited $CF_4$ molecules emit photons from the far UV to the visible~\cite{pansky}.
 This results in an intolerable level of afterpulsing in $Ar/CF_4$ gases even at moderate
gas gains.
The advantage of the enhanced drift velocity of $CF_4$ for high-rate
applications have been realized by the addition of
one of the common quenchers (e.g.\ $CO_2$, $CH_4$) to $CF_4$ or to $Ar/CF_4$.
This can also `cool' electrons to the extent that attachment does not occur.

\subsection{ Lessons learned from detector operation at high ionization densities.}

Accelerated aging tests, often carried out with radioactive sources or $X$-rays 
to emulate the long-term lifetime properties of the
detector, can be extrapolated to the real experimental conditions directly only if the aging rate
depends on the total collected charge and not upon the current density, particle rate, or gas gain at which the dose was accumulated.
In reality, many laboratory studies have demonstrated severe gain
losses at lower charge accumulation rates, other conditions being held 
constant~\cite{cern1,opensh1,kotthaus,algeri,juri,bouclier}.
  The lower polymerization rate for higher current densities
is attributed to the onset of space charge effects, which
reduce the electron energies in the avalanche, thus
decreasing the density of ions and radicals in the avalanche plasma.

 The new generation of high-rate detectors of the LHC era has not only to cope with
high dose rates, but also has to survive 
in a hostile presence of heavily ionizing particles with an average energy deposition 
10-1000 times larger than for MIP's.
An exposure of `large-scale' gaseous detectors over the full area in such a 
harsh radiation environment at high ionization densities ($>100~nA/cm$) 
can result in greatly enhanced polymer formation:
an abundance of aggresive radicals
will diffuse for rather long times ($\sim$~hours) within the irradiated chamber and
will react with other avalanche-produced polymer fragments.
According to this naive picture, this mechanism could significantly accelerate polymerization in large systems,
whereas in small-scale laboratory tests 
the aging rate typically decreases with increasing gas flow~\cite{kadyk,opensh1,opensh2,atlas2}.
 Furthermore, polymer deposition in wire chambers likely starts from localized spots 
and then can spread over the entire irradiated region. 
Since in large, mass-produced systems an extremely high quality for
electrode surfaces and cleanliness of the gas systems are hard to reach, 
any imperfections -- in the presence of high currents --
could easily trigger sparks, discharges or Malter currents, which will
in turn dramatically increase the polymerization rate.

Recent systematic research clearly demonstrates that 
the initial stage of radiation tests 
usually performed in the laboratory may not offer the full information needed to
estimate the lifetime of the real detector.
 Strong dependence of aging performance upon
{\bf{ size of the irradiated area, current density, high voltage 
(gas gain), irradiation rate, particle type and energy}} have been observed in
high-rate environments. 

 Severe anode and cathode aging effects were found in prototype honeycomb drift
tubes operated with $CF_4/CH_4$ (80:20) and $Ar/CF_4/CH_4$ (74:20:6) mixtures in the high-rate
HERA-B environment (secondaries from interactions of 920~GeV proton with target nucleus)
after a few $mC/cm$ of accumulated charge~\cite{bohm,padilla,kolanoski,hohlmann2,stegmann}.
This effect was surprising since chambers had previously been proven to be immune
to very large $X$-ray doses up to 5~$C/cm$.
An extensive $R\&D$ program, carried out at 10 different 
radiation facilities to resemble HERA-B conditions, revealed 
that $X$-rays or electrons were not able to trigger Malter currents, 
while in the large-area modules, irradiated
with hadrons above a certain energy, Malter effect appeared very rapidly.
The aging effects in these honeycomb drift tubes were traced to
a combination of several problems; a solution which uses gold-coated cathode foils 
and a $Ar/CF_4/CO_2$ (65:30:5) mixture was found to survive in a high-rate 
hadronic environment up to $\sim~1~\frac{C}{cm~wire}$~\cite{padilla}.

 The aging performance of the HERA-B muon proportional chambers 
has been studied with $Ar/CF_4/CH_4$, $CF_4/CH_4$, and $Ar/CF_4/CO_2$ 
mixtures in a variety of conditions~\cite{titov1}-\cite{titov4}.
 The aging rate in $Ar/CF_4/CH_4$
was found to be more than two orders of magnitude higher 
in hadronic beams than in the laboratory studies with radioactive sources.
In addition, strong dependences of the aging properties on high voltage and 
progressive deterioration of the gas gain in the direction of the serial gas flow
have been observed for large-scale prototypes irradiated with
$Ar/CF_4/CH_4$ (67:30:3) in the harsh HERA-B environment.
Aging effects increasing in the direction of the serial gas flow (even
outside the irradiation zone) have been also reported for a $Ar/CO_2/C_2H_2F_4$ mixture~\cite{lobachev}.
A strong dependence of the aging properties on high voltage, irradiation rate, length of
irradiation and gas flow rate has been also observed
in the ATLAS muon drift tubes operated with $Ar/CH_4/N_2/CO_2$ 
(94:3:2:1)~\cite{pash,atlas2,kollef2,atlas}.
 Here, the systematic $R\&D$ studies have shown a nearly exponential dependence of chamber lifetime
on high voltage and on the counting rate within the experimentally investigated parameter range.
Unfortunately, the high voltage 
is not the physical quantity directly responsible for aging in wire chambers, therefore,
these aging effects could be classified 
as depending on the gas gain and/or current density,
which are related to the density of ions and radicals in the avalanche plasma.

 Dependence of polymer formation on the energy input is well established in plasma 
polymerization. 
Nearly all organic compounds 
regardless of their chemical nature can be polymerized under certain conditions.
The structure of plasma polymers formed from the same monomer is highly dependent on the 
actual conditions of polymerization: the energy input level, the size 
(cross-sectional area) of a tube and even on the position within the reactor~\cite{yasuda}.
The experimental dependence of chamber lifetime on 
ionization density, gas gain and irradiation rate, which are also related to the
total dissipated energy in the detector from ionizing particles, 
indicates that the aging behavior 
can not be solely explained on the basis of the molecule ratios in the mixture (gas composition),
without taking into account the actual operating conditions.
These results illustrate the need for studying the aging performance of a detector under
conditions as close as possible to the real environment.

The dependence of the detector lifetime on the size of the irradiated area, in 
particular,
and the increase of the aging rate in the direction of the serial gas flow
means that aging should be viewed as a non-local and intensity-dependent phenomenon.
These observations seem to be the most critical when trying to extrapolate the aging
behavior from laboratory tests to large-scale detectors.
 Some of the long-lived aggressive radicals may diffuse in the direction
of the gas flow and react with other avalanche generated contaminants, thus 
enhancing aging effects with increasing usage of the gas.
Here it is important to note
that due to the increased aging in the direction of the gas flow it is worthwhile to 
avoid gas distribution systems that supply many chambers by a serial flow.

\begin{figure}[bth]
\setlength{\unitlength}{1mm}
 \begin{picture}(100,40)
\put(-5.0,-42.0){\includegraphics{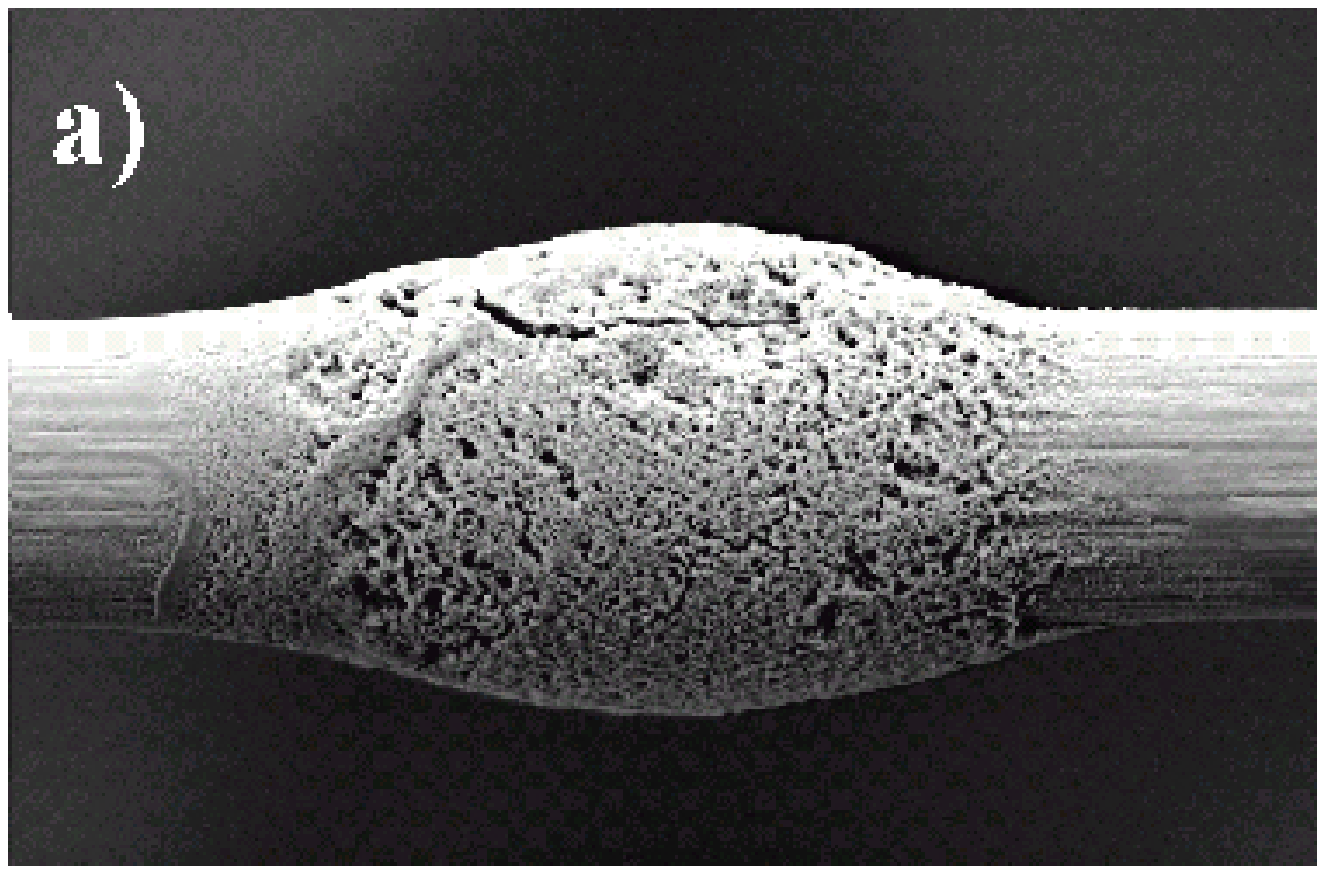}}
\put(85.0,-42.0){\includegraphics{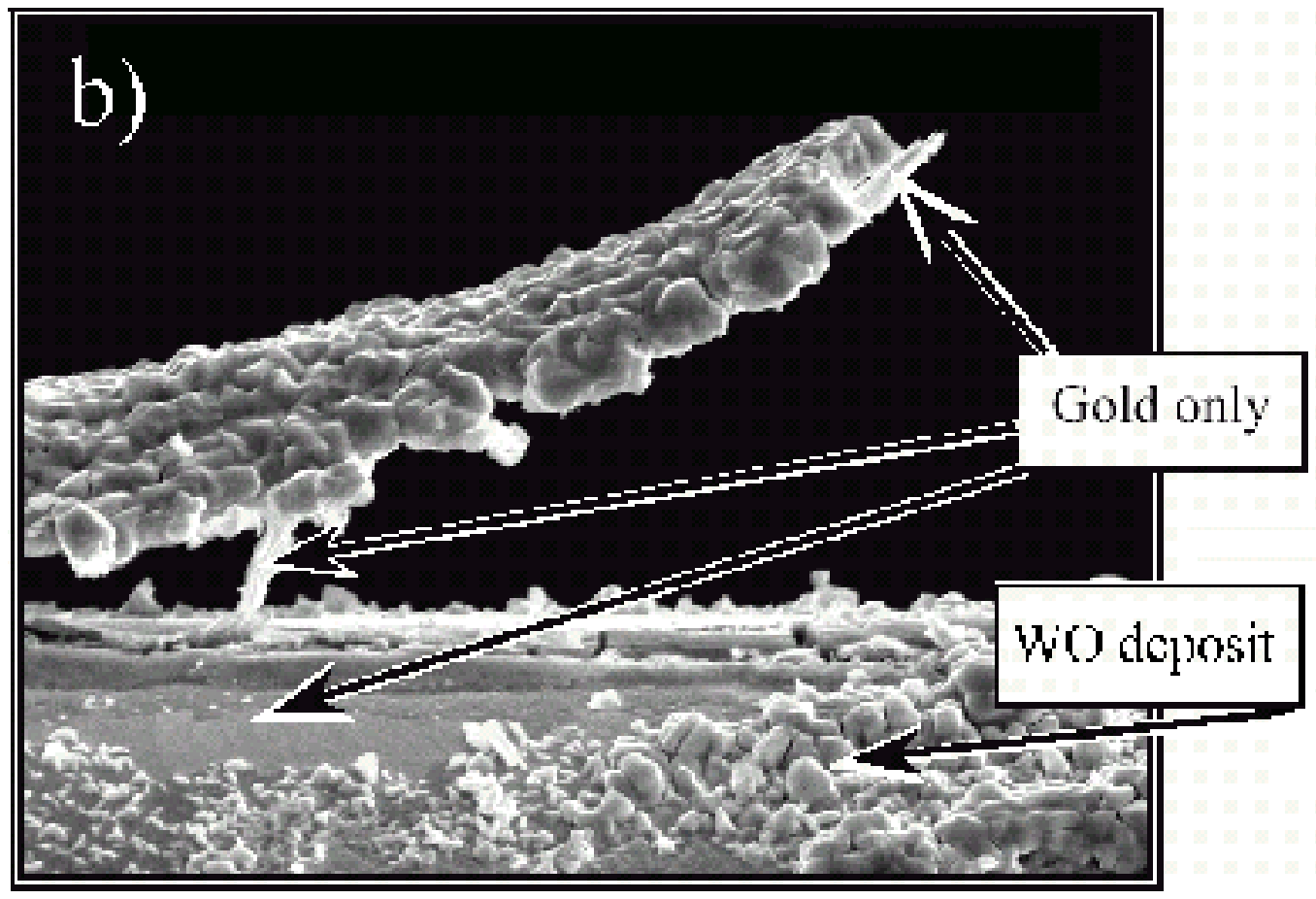}}
 \end{picture}
 \caption{ Various manifestations of damage to the gold-plating of wires:
a) amorphous $WO$ deposits~[147]; 
b) long piece of gold peeled away from the wire surface~[148].
 \label{damage}} 
\end{figure}

As reported above, polymer deposition can occur in $CF_4/CH_4$ mixtures, as suggested
by plasma chemistry. Similarly,
the aging properties of $Ar(Xe)/CF_4/CO_2$ gases, which by analogy with
plasma experiments should have excellent etching properties,
have been widely investigated over the last years.
 Using $Ar/CF_4/CO_2$ mixtures under optimal operating conditions,
no observable drop in gain due to polymerization has been found
for the HERA-B honeycomb drift tubes up to 1.5~$C/cm$~\cite{padilla},
HERA-B aluminum proportional chambers up to 0.7~$C/cm$~\cite{titov1},
CMS cathode strip chambers up to 0.4~$C/cm$~\cite{prokofiev}, 13~$C/cm$~\cite{gavrilov},
LHCb multi-wire proportional chambers up to 0.25~$C/cm$~\cite{souvorov}, 
COMPASS straw tubes up to 1.1~$C/cm$~\cite{dunnweber} and 
HERMES drift tubes up to 9~$C/cm$~\cite{hermes}. 
Moreover, honeycomb drift tubes which were initially aged with $Ar/CF_4/CH_4$ 
were afterwards successfully recovered in $Ar/CF_4/CO_2$~\cite{padilla}.
However, an analysis of the cathode surfaces at the
end of operation with $Ar/CF_4/CO_2$ revealed in some cases the presence of
fluorine-based deposits on the cathodes, which fortunately  
did not result in self-sustained currents~\cite{gavrilov,titov1,prokofiev}.
Since dissociative products of $CF_4$ react violently with many materials and
the resultant polymer films at cathodes could  provoke aging effects
one should seriously consider using materials in high-rate detectors
which are very robust to $CF_4$; gold-plated electrodes~\cite{padilla} or straw cathode 
materials~\cite{rromaniouk} fulfill this requirement.

A further advantage of $CF_4$-based mixtures is additional resistance
against $Si$-polymerization. 
This suggestion is based on experiences from plasma polymerization, where 
discharges of $CF_4/H_2$ are successfully
used for $Si O_2$ etching, while $CF_4/O_2$ plasmas selectively etches $Si$~\cite{kushner}.
 Extensive studies performed for the ATLAS straws partially 
confirm this hypothesis: $Si$-deposits have not been observed in irradiation area
for large current densities ($>1~\mu A/cm$).
On the other hand, $SiO/SiO_2$-deposits were found at the edges 
and even outside of the irradiated area.
The resulting balance between $Si$ polymerization and 
$CF_4$ etching processes was found to be very sensitive to the $Si$-source intensity and 
ionization density~\cite{rromaniouk}. 


In the most recent investigations at
extremely high current densities ($\sim 1-5 \mu A$/$cm$)
in $Ar(Xe)/CF_4/CO_2$ mixtures
a new `aging phenomenon' has appeared - the damage of the gold-plating of wires
in straws and honeycomb drift tubes.
Several years of intensive research at CERN with straw tubes
under different conditions and for different types of wires indicate that the
main components responsible for gold wire damage in $Xe/CF_4/CO_2$ (70:20:10) 
are harmful radicals, products of $CF_4$ disintegration, in connection with $H_2O$. 
(It was suggested that $HF$ acid could be
responsible for destruction of gold-plating and formation of $WO$ deposits).
No wire damage effects have been observed for water concentrations 
below 0.1~$\%$ up to 20~$C/cm$~!~\cite{rromaniouk}.
 Dedicated studies with straw tubes performed at PNPI with
$Xe/CF_4/CO_2$ (70:20:10) and $Ar/CF_4/CO_2$ (60:30:10) mixtures
demonstrated that under high dose rates 
the gold-plating of the wire was cracked, the wire diameter increased
and a large amount of oxygen
was observed on the tungsten in gold cracks~\cite{krivchitch1}-\cite{krivchitch3}.
Similar effects have been observed for wires
irradiated in a $Ar/CO_2/C_2H_2F_4$ mixture~\cite{lobachev}.
The authors propose a model to explain the results - an anode wire `swelling' mechanism,
where the forces causing the damage to the wire surface develop under the gold
layer of the wire~\cite{krivchitch2}.
Fig.~\ref{damage} shows examples of a variety of wires with damaged gold-plating 
from~\cite{rromaniouk,krivchitch1}.
In contrary to the experience with straws, 
in one test with honeycomb drift tubes irradiated with $Ar/CF_4/CO_2$ (65:30:5)
the destruction of gold coating and even rupture of anode wires have been observed
only for water concentration below 50~$ppm$,
while for $H_2O > 400~ppm$, gold wire damage effects were avoided~\cite{schreiner1,schreiner2}.
Further studies still remain to be done to fully understand the exact mechanism of gold wire damage
during operation at high ionization densities.
Although no $F$-based deposits were observed on the anode wires in any of the tests,
the chemically reactive dissociative products of $CF_4$ 
most probably initiate the destruction processes of gold-plating.

 Studies of straw proportional tubes with $Xe/CF_4/CO_2$ mixture 
revealed another phenomenon, which might degrade detector performance
in high rate experiments.
The gas composition was found to be modified in the avalanche plasma
of a strongly irradiated straw, presumably due to the production of 
some long-lived and highly electro-negative species~\cite{electr1}.
 These electro-negative radicals could be also responsible for the so-called `transient aging effect'
observed at high-rates in
$Xe/CF_4/CO_2$~\cite{electr2}, $Ar/CF_4/CO_2$~\cite{schreiner1} 
and $Ar/CF_4/CH_4$~\cite{titov1}. A `transient aging effect' is a 
temporary gain reduction,
which can be restored by an appropriate increase of the gas flow.
The very high aggressiveness of dissociative products of $CF_4$ and the 
dynamic modification of the gas composition requires more detailed studies to evaluate
the possible consequence of these effects on the long-term performance and stability 
of large-area gaseous detectors.
In view of the aging results described here, one can see that the presence of large
amounts of $CF_4$ in the mixture does not necessarily ensure good aging properties
automatically.

 The challenge to avoid aging in the new generation of high-rate experiments requires
not only a very careful choice of all detector materials, but also forces the gas systems
to be of previously unnecessary quality and cleanliness.
However, a real system will always contain some degree of imperfection and pollution
-- despite all precautions. 
It has to be stressed here that
in closed-loop recirculation systems, 
which are required for detectors operated with expensive gases ($Xe,CF_4$)
all impurities and reactive radical fragments will remain 
in the gas until they are removed by a purification system or deposited elsewhere.
Therefore, for the construction of large-area gaseous detectors
the maximal cleanliness for all
processes and quality checks for all system parts are of primary importance.
Examples of `clean' gas systems currently used for high-rate detectors 
are presented in~\cite{hohlmann,dreis}.
Certainly, the definition of the word `clean' has changed considerably since
 the 1986 workshop.

\section{ Summary }

Aging phenomena
obviously constitute one of the most complex and serious problems which could limit
the use of gaseous detectors in unprecedently severe radiation environments.
The operation in high-intensity experiments of the LHC era demand not only an extraordinary
radiation hardness of construction materials and gas mixtures, but also  
appropriate assembly procedures and quality checks during detector construction and testing.
 Since the 1986 workshop, 
considerable progress has been made on the understanding of general principles 
which might help to prevent or at least to suppress the aging rate to an acceptable level.
However, a quantitative description of the aging processes, which would require a
detailed knowledge of the reaction cross sections of all chemical species
in the avalanche plasma, is still not available.

After the many years of intensive research and development of radiation-hard 
gaseous detectors, 
an impressive variety of experimental data has been accumulated.
The radiation hardness and outgassing properties of the various
materials used for the construction of detectors
and gas systems are among the most crucial items affecting the lifetime
of gaseous detector. 
However, the observed dependences of aging performance on the 
nature and purity of the gas mixture, different additives and trace contaminants,
construction materials, gas flow, size of the irradiated area, 
irradiation intensity, ionization density,
high voltage, particle type and energy,
make quantitative comparisons of aging properties under very different conditions
very difficult. 
Consequently, this data can serve only as a guideline before the start of long-term studies
under conditions as close as possible to the real environment of the experiment.
 Such radiation tests should include an extended study of `large-scale' final prototype
chambers, exposed over the full area to a realistic radiation profile 
(particle type and energy, ionization density, irradiation rate). 
It is of primary importance to vary the operating parameters systematically in order
to investigate their possible influence on the aging performance.
In order to exclude statistical fluctuations of unknown nature and to provide
a reliable estimate for the detector lifetime, the radiation tests should
be carried out with several detectors irradiated under identical conditions.

This paper is based on the results reported at the DESY workshop, 
and also briefly discusses other experience with gaseous detectors relevant to the present aging problems.
Transparencies and videos of presentations
from the `International Workshop on Aging Phenomena in Gaseous Detectors' (DESY, Hamburg)
are available at the workshop's web-page (http://www.desy.de/agingworkshop).  
 The proceedings of the workshop will be published in a special volume of Nuclear 
Instruments and Methods: Section A.

\section{Acknowledgment}

We are especially grateful to Andreas Schwarz (DESY), Jaroslav Va'vra (SLAC)
and Mar Capeans (CERN) for many helpful discussions and careful reading
of this manuscript.

\end{document}